***Camilla Mazzucato**: Department of Cross-cultural and Regional Studies, University of Copenhagen (DK), camimazz@hum.ku.dk,orcid: 0000-0002-4168-0773 -* ***Corresponding author***
***Michele Coscia**: IT University of Copenhagen (DK), mcos@itu.dk, orcid: 0000-0001-5984-5137*
***Ayça Küçükakdağ Doğu**: Department of Biological Sciences, Middle East Technical University, Ankara (TR), ayca.kucukakdag@metu.edu.tr, orcid: 0000-0001-6208-4092*
***Scott Haddow**: Department of Cross-cultural and Regional Studies, University of Copenhagen (DK), scott.haddow@hum.ku.dk, orcid: 0000-0002-3970-7447*
***Muhammed Sıddık Kılıç**: Department of Health Informatics, Graduate School of Informatics, Middle East Technical University, Ankara (TR), siddik.kilic@metu.edu.tr orcid: 0009-0009-2341-1471*
***Eren Yüncü**: Department of Biological Sciences, Middle East Technical University, Ankara (TR), eren26285@gmail.com*
***Mehmet Somel:** Department of Biological Sciences, Middle East Technical University, Ankara (TR), somel.mehmet@gmail.com, orcid: 0000-0002-3138-1307*

# "*A network of mutualities of being*": socio-material archaeological networks and biological ties at Çatalhöyük.

## Abstract

Recent advances in archaeogenomics have granted access to previously unavailable biological information with the potential to further our understanding of past social dynamics at a range of scales. However, to properly integrate these data within archaeological narratives, new methodological and theoretical tools are required. Effort must be put into finding new methods for weaving together different datasets where material culture and archaeogenomic data are both constitutive elements. This is true on a small scale, when we study relationships at the individual level, and at a larger scale when we deal with social and population dynamics. Specifically, in the study of kinship systems it is essential to contextualize and make sense of biological relatedness through social relations, which, in archaeology, is achieved by using material culture as a proxy. In this paper we propose a Network Science framework to integrate archaeogenomic data and material culture at an intrasite scale to study biological relatedness and social organization at the Neolithic site of Çatalhöyük. Methodologically, we propose the use of *network variance* to investigate the concentration of biological relatedness and material culture within networks of houses. This approach allowed us to observe how material culture similarity between buildings gives valuable information on potential biological relationships between individuals and how biogenetic ties concentrate at specific localities on site.

**Keywords**: Southwest Asia, Çatalhöyük, Neolithic, aDNA, Kinship, Network Science, Network Variance.

## 1.Introduction

Recent developments in aDNA extraction and sequencing (NGS) have unlocked unprecedented access to ancient biological relationships between individuals. Many studies have used these datasets to attempt the reconstruction of past kinship systems and patterns of social relations (e.g. Blöcher et al., 2023, Fowler at al., 2022; Ning et al., 2021; Rivollat et al., 2023; Sjögren et al., 2020; Sikora et al., 2017; Somel et al., 2023; Yaka et al., 2021). Archaeologists are increasingly engaged in debates surrounding the integration and interpretation of archaeogenomic data within archaeological research and the affordances of archaeogenetics in the reconstruction of past conceptions of kinship and relatedness (Brück, 2021,2023; Brück and Frieman 2021; Frieman, 2021, 2023; Frieman and Hofman, 2019; Furholt 2021; Hofmann et al., 2021; Risch et al., 2023; Thelen, 2023). While archaeogenetic research provides archaeologists with a valuable account of the biological relations of past individuals, our ability to conceive the contours and complexities of prehistoric kinship systems remains limited. In a recent article, Joanna Brück (2021) warns archaeologists of the risk of "biological essentialism" and of reducing kinship to solely biological ties. She provides a review of anthropological perspectives on kinship, stressing the role of "cultural selection", which is always "built into kinship systems'" (Brück, 2021:2). In the same article she states that, while aDNA analysis has the potential to improve our understanding of the past, the results are often difficult to understand and our interpretations often problematic; archaeological data (e.g. material culture, bioarchaeological data) should be used to put genetic results in context. In a response to the above-mentioned debate article, Rachel Crellin states: "*We need to create heterogeneous assemblages of data to explore kinship critically. aDNA can, of course, be part of this, but it should be one component of many. The archaeological record is undoubtedly a rich place in which to explore the many, varied and everchanging forms of kinship"* (Crellin, 2021:2).

In this paper, we assemble a heterogeneous dataset composed of material culture and genetic data with the aim of enriching our understanding of the social organization and modes of relatedness at the Neolithic site of Çatalhöyük. Specifically, we use methods borrowed from Network Science to analyze a complex dataset comprising a selection of material culture data recorded by the Çatalhöyük Research Project (ÇRP) between 1993 and 2017, and biological relatedness of individuals deriving from aDNA analysis. By combining material culture and genetic data, we seek to investigate how biological relations between individuals correlate with variation in material culture and how this can shed light on the construction of biosocial relations at Çatalhöyük. In doing so we seek to investigate how relatedness is materialized and to highlight the importance of materiality in constructing and reproducing kinship bonds.

This is done through the integration of archaeogenomic data extracted from individuals buried underneath the floors of houses and assemblages of co-occurring material practices between the same houses.

Since at least the 1980s, social anthropologists have critiqued the conflation of different ideas of kinship with a Eurocentric view of kinship relations as intrinsically related to biological descent and of a separation between social processes and biology (e.g. Collier and Yanagisako, 1987; MacCormack and Strathern, 1980; Schneider, 1984). Similarly, the assumption that kinship represents an isolated and static societal domain has been refuted (e.g. Carsten, 2000, 2004, 2019; Sahlins, 2013; Strathern 2014). On the contrary, the creatively constructed, dynamically negotiated and historically situated nature of kinship bonds is now largely acknowledged and while biological relations may be differentially incorporated within the various "cultures of relatedness" (Carsten, 2000), the exchange of genetic material is just one of many means for creating bonds between individuals. Numerous ethnographic examples show that elements of cultural selections are present even when biological relatedness is acknowledged as creating kinship ties. A striking example is the one of the Mukkuvar (Kerala, South India) where mothers and fathers contribute different substances in the procreation, and they are therefore differently connected to their children (Busby, 1997, 1997b). Similarly, societies that follow a purely unilineal descent, matrilineal or patrilineal, decide to completely disregard biological motherhood or fatherhood in the construction of kinship relationships (Johnson 2016; Sahlins, 2013).

The fluid and variable nature of kinship systems has been extensively highlighted by scholars who study the problematic intersection of indigenous definitions of relatedness and belonging, and the biological identification of indigenous populations via genetic studies alone. These scholars have written extensively on the conflicting realities of genomic knowledge and indigenous identity practices (Abel and Frieman, 2023; Abel and Schroeder 2020; TallBear, 2013a, 2013b, 2018). Indigenous and feminist scholars like Kim TallBear (2018) or Paulla Ebron and Anna Tsing (2017) have spoken of the colonial violence related to the imposition of Western ontological categories of heteronormative families on indigenous populations in which the processes of kin-making are incredibly diverse (Brück and Frieman, 2021).

In *What Kinship Is – And Is Not,* Marshall Sahlins argues that a kinship system is a "*manifold of intersubjective participations, which is to say, a network of mutualities of being*" (Sahlins, 2013:20). In this way, he defines kinship as culturally and historically variable and socially constructed. With this definition, Sahlins captures the mutable, inclusive nature of kinship ties, which encompass a variety of processes and mechanisms (Carsten, 2019). A large body of ethnographic literature reveals the "processual" nature of kinship configurations; kin relationships, in these cases, are not a *fait accompli* but relatedness is acquired through specific processes and their configurations may change through time (Bamford, 2009; Carsten, 1995, 2004; Johnston, 2021; Weismantel, 1995). Such kinship relations frequently center on food sharing and co-habitation. For example, Nurit Bird-David, in her work with small communities of South Indian foragers, argues that kinship develops within these groups as an immediate, intimate “pluripresence” of diverse relatives where it is impossible to disentangle one kinship relation from another (Bird-David, 2017). Janet Carsten (1995, 1997), working among the Malays on the island of Langkawi, observes the fluid nature of kinship ties and the importance of consuming food in the same houses and participating in activities in order to form kinship bonds. Kinship bonds between humans and non-humans are also constructed and maintained through co-living and through relentless social practice. Indeed, as stressed by Brück and Frieman (2021), archaeologists are in a privileged position to observe the way social practices contribute to the creation and endurance of kin relations. Moreover, a primary focus on materiality makes archaeologists especially aware of the role played by objects and the built environment in the creation and maintenance of kinship bonds. For example, Johnston (2020) examines the role played by specific hoards and objects in Bronze Age Britain and Ireland. In Johnston’s view, such assemblages enabled the formation of relations through their long biographies that bound people together across time and place (Johnston 2020). These assemblages were instrumental in producing and sustaining kinship relations and facilitating “kinwork” through the gathering of humans and non-humans in a specific location (Johnston 2020). Carsten (2019) notices how kinship qualities “have the tendency to attach to stuff” (2019:39) and they are entangled in the creation and maintenance of kinship bonds. A rich body of ethnographic case studies demonstrates the fluid manner in which human groups generate patterns of meaningful relations and the broad spectrum of entities involved (see Crellin, 2021; Crellin at al., 2020; Sahlins 2013). Such configurations of relations may include non-human animals, inanimate objects, or spirits and do not require exchange of genetic material. Famously, Donna Haraway unpacked the intimate and generative relation that links humans to their “significant other” pets in kinship associations (Haraway, 2006). Ethnographic research conducted amongst the Nayaka, a group of immediate-return foragers of South India, showed how non-human animals were ubiquitous presences incorporated within human “pluralist communities” as “close relatives” and family (Bird-David, 2006:47). Countless ethnographic examples testify to the complexity of human-animal relationships and in general to the entwined relation between humans and non-human entities that include sharing of kinship bonds and processes of identification (e.g. Descola 2013; Hallowell, 2012; Viveiros de Castro, 2012). Kin relationships between non-human entities and humans can extend to land and places which in ethnographic accounts appear to be constitutive of kinship bonds (see Bamford, 1998, 2007, 2009; Basso, 1996; Leach, 2003, 2019; Morphy, 1995; Telban, 2019).

## 2.Background

### 2.1 The site

Çatalhöyük is a large Neolithic tell site located in the Konya Plain in Central Anatolia. The site was excavated by James Mellaart between 1961 and 1965 and by Ian Hodder between 1993 and 2017 (Hodder and Tsoraki, 2021; Mellaart, 1967). The Neolithic occupation of the settlement is situated on the 13.5ha East Mound and dates between 7100 and 5950 cal BCE (Bayliss et al., 2015, 2022; Marciniak et al., 2015a) (Figure 1). The occupation of the East Mound is characterized by sequences of mudbrick buildings constructed one on top of the other and, at intervals, separated by external spaces such as middens and penning yards. Çatalhöyük shows no signs of deliberate planning; instead, the site seems to have developed an organic, modular arrangement through the repetition of similar structures. Most scholars agree that these structures are all similar residential units which are spatially distinguishable from one another and whose modular proximity form the dense agglomeration of the East Mound (Hodder, 2013, 2014). At the site, there is ample evidence of small-scale, house-based organization of production and food consumption (Bogaard et al, 2017). Houses appear to be all equal entities and to be the main center of the productive and ritual life at the site. Nevertheless, differences between buildings have been detected at the site since the first excavations; this is especially evident in the variable frequency of burials within buildings at the site (Haddow et al. 2021). Intramural, subfloor primary burials, including all age and sex groups, are the dominant burial practice at Çatalhöyük; previous researchers have assumed that co-burials represent family (i.e. biological) groupings, but in fact we know very little about how these people are related and why they were selected for internment. Recent work on the burial assemblage at Çatalhöyük has revealed complex patterns of post-mortem body treatment. Skeletons show varied degrees of completeness, flexion and preservation and some of them appear to have been wrapped in animal hides, textile, cordage or matting. Such variations are potentially related to complex, multi-stage burial practices involving a delay between death and final burial for some community members (Haddow et al., 2021; Haddow and Knüsel, 2017). Furthermore, histotaphonomic research (Haddow et al., 2023) suggests that bodies buried underneath the floor of the same building have diverse post-mortem biographies and that they reached their final place of interment after being kept above ground for different lengths of time and under different conditions. Indeed, it has been suggested that some of the bodies might have been kept circulating in buildings or external areas as desiccated or skeletonized bundles (Haddow et al., 2021, 2023). The social logic that underscores the formation of burial assemblages at Çatalhöyük is very complex and remains poorly understood. Regarding the social organization of the site, different theories have been proposed since the site was discovered by James Mellaart in 1958. Some scholars emphasize the role of geographical proximity and clustered neighborhoods as the main principle of social organization as at other Neolithic sites (e.g Aşıklı Höyük) (Düring, 2007; Düring and Marciniak, 2006). Kuijt (2018) suggests that individual buildings should be more appropriately regarded as rooms serving different functions within larger complexes. These complexes are thought to have been used by an extended/multi-family household made up of different families/components kept together by affiliations distributed spatially in building clusters.

In this regard, it should be stressed that recent excavations in the South and North Areas have revealed a highly complex scenario concerning localized continuity (overlapping buildings) and larger neighborhood continuity (Bayliss et al., 2022; Farid et al.,2023). It now seems clear that there were frequent breaks in the sequence of reconstructions of stacks of buildings during the occupation of the North Area and that open space accounted for a considerable proportion of the North Area at any one time (Bayliss et al. 2023). We now observe a more dynamic and dispersed type of habitation at the site that does not seem to fit the clear-cut picture of a clustered neighborhood arrangement. Other scholars highlight the role of supra-household groups: Mills (2014) for instance, sees Çatalhöyük's social fabric arranged in "flexible networks" of cross-cutting social groupings (e.g. religious sodalities) that overlap to form a tightly knit society (Hodder 2014a). Hodder and Pels (2010) suggest a process of differentiation between ordinary houses and "history houses" through the accumulation of social memory. History houses were extended entities that functioned as "foci for groups of houses"; the members of the extended house were very likely sharing a wide number of activities e.g. working together, eating together and being part of the same rituals that were, perhaps, happening in the history house (Haddow and Pels, 2010). As such, history houses represent a series of "cross-cutting" dependencies that formed a deeply interconnected "social mosaic" (Hodder 2014a:155). Most recently, Hodder (2022) points to a system that in the Early and Middle periods (Table 1) was characterized by a mixed focus on grouping of proximate buildings and sets of affiliations and alliances that were cross-cutting the geography of the site. Within such a system, "history houses" (or elaborate houses) had a "nodal" function" as aggregators of specific groups, sodalities or lineages (Hodder 2022:13). By the Late occupation (post-6500 cal. BCE), this balanced system is replaced by a much more dispersed and fragmented arrangement of occupation at the site (Hodder, 2022; Marciniak et al. 2015b). Buildings in this period appear to be more autonomous and the landscape around the site appears to be exploited in a much more extensive way. These changes encompass all aspects of Çatalhöyük society and are major and dramatic transformations (Czerniak and Marciniak 2022; Hodder, 2013, 2014a; Marciniak, 2019; Marciniak et al., 2015b).

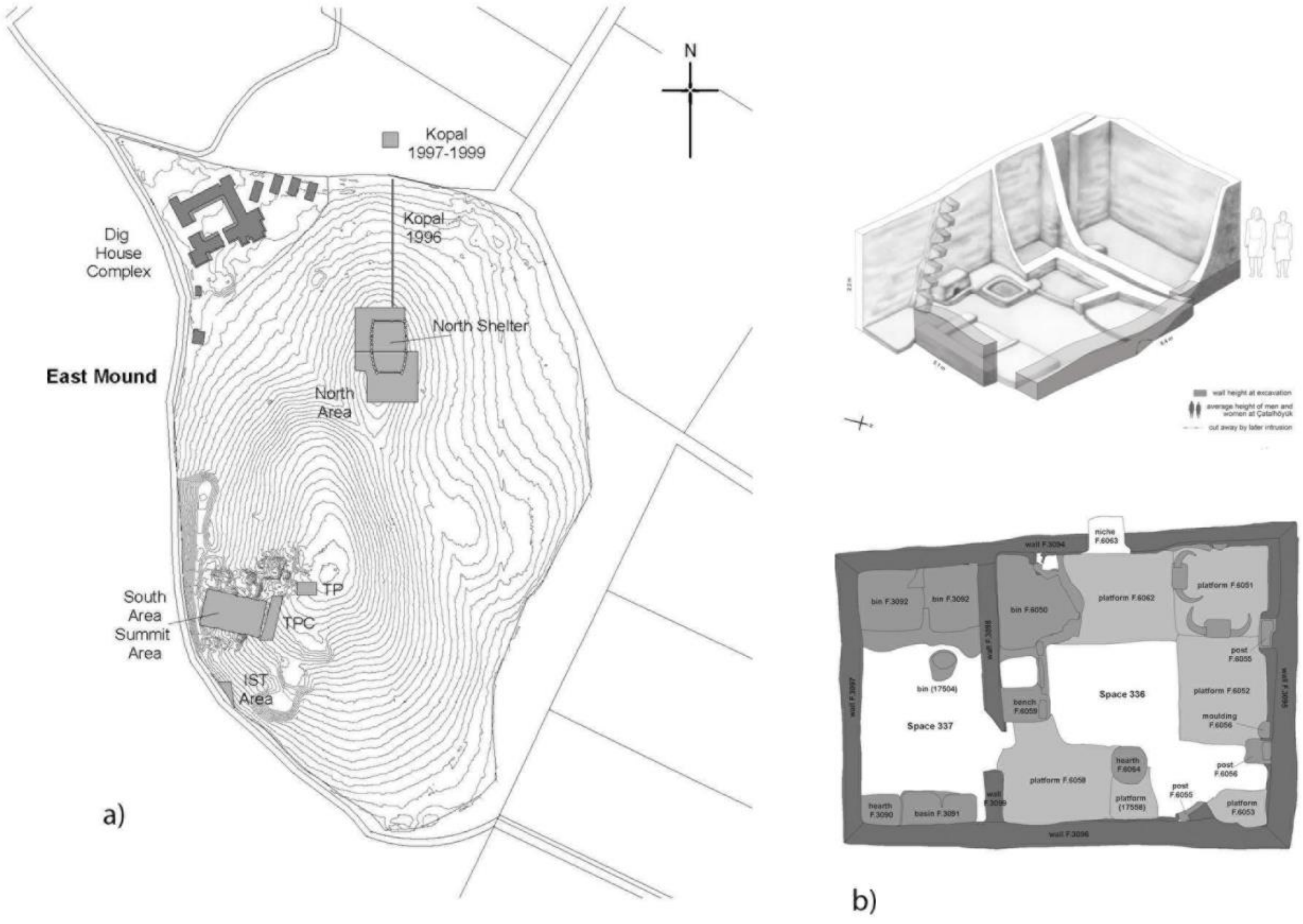

**Figure 1**: a) Çatalhöyük East Mound with highlighted excavated areas. b) Reconstruction of Building 58 and plan of Building 77 (Courtesy ÇRP and Killackey Illustration).

| **Çatalhöyük East levels and temporal groupings** | | | | | | |
|---|---|---|---|---|---|---|
| **Temporal groupings of levels** | **South** | **North** | **TP, TPC, GDN** | **IST** | **Mellaart** | **Cal BCE** |
| **Final** | | | TP.Q-R | | **I** | 6300-5950 BC |
| | | | TP.O-P | | **II** | |
| **Late** | South.T | North.H-J | TP.N | IST | **III** | 6500-6300 BC |
| | South.S | | TP.M | | | |
| | South.R | | TP.L | | **IV** | |
| | South.Q | | | | | |
| | South.Pb | | | | **V** | |
| | South.Pa | | | | | |
| **Middle** | South.O | North.F-G | | | **VIA** | 6700-6500 BC |
| | South.N | | | | **VIB** | |
| | South.M | | | | **VII** | |
| **Early** | South.L | | | | **VIII** | 7100-6700 BC |
| | South.K | | | | **IX** | |
| | South.J | | | | **X** | |
| | South.I | | | | **XI** | |
| | South.H | | | | **XII** | |
| | South.G | | | | | |

**Table 1**: Çatalhöyük East phases and macro-temporal groupings Early, Middle, Late and final).

## 2.2 Archaeogenomic research at Çatalhöyük

Archaeogenomic analysis at Çatalhöyük has recently been conducted by the NEOGENE Project team, which genetically screened the skeletal remains of 381 individuals dated to the Neolithic period (Yüncü et al., 2024; Yaka et al., 2019). As a result, the team was able to infer genetic relatedness for 131 individuals (Yüncü et al., 2024). The analyses identified different degrees of relatedness between pairs of individuals based on the amount of genetic material shared. First degree-related individuals share about 50% of their genetic material, representing, for example, a parent-child, or sibling relationship. Individuals that share around 25% of their DNA are considered 2nd-degree relations (avuncular pairs, such as aunt-nephew, half-siblings, and grandparent-grandchild pairs), while 3rd-degree relations share approximately 12% of their genetic material (first cousins, great avuncular pairs, etc.) (see Table 2). The best genetic preservation has been found in sub-adult skeletons, which form the majority of the genetic dataset. The striking difference in genetic preservation between adults and sub-adults at Çatalhöyük strongly suggests differential funerary treatments between age groups (see Haddow et al., 2022, 2023).

Initial archaeogenomic research undertaken on a small sample of individuals at Çatalhöyük and other Neolithic Anatolian sites has highlighted a temporal trend in the degree of biological relatedness in co-burials within buildings. Yaka and colleagues (2021) reported that biological relatedness appeared to decrease in importance as a selection principle for individuals buried within the same building from Aceramic Neolithic sites, Aşıklı Höyük and Boncuklu Höyük, (9th-8th millennia) to Early Ceramic sites in the 7th millennium BCE, Çatalhöyük and Barcın Höyük. Recent results have revealed that a similar trend unfolds through the Çatalhöyük sequence, with Early period buildings more likely to contain genetically related individuals than those in the Late period (Yüncü et al., 2024). Furthermore, a pattern of genetic descent focused on maternal lineages is also observed (*Ibid*), wherein all individuals frequently share their mitochondrial DNA with other individuals buried within the same house; on the contrary, males buried in the same building do not frequently share their Y chromosome. This persistence of maternal genetic descent within houses is potentially consistent with a matrilocal or mixed matrilocal residence model (Yüncü et al., 2024), in the sense of adult females remaining connected with the building but adult males tending to move away. We note that these findings are in opposition to a previous smaller study focusing on mitochondrial aDNA extracted from skeletons buried underneath the floors of three adjacent Çatalhöyük buildings, which was unable to identify genetic similarities between maternal lineages among co-burials (Chyleński et al., 2019). However, that study was based on 10 individuals only.

| Biological relatedness | |
|---|---|
| 1st degree | Parent-child or siblings (exactly 50% DNA shared between parent-child, on average 50% shared between siblings) |
| 2nd degree | Grandparent and grandchild, half siblings, aunt/uncle and niece/nephew of an individual (25% DNA shared on average) |
| 3rd degree | First cousin, great-grandparent, great-aunt/great-uncle, great-niece/great-nephew, great-grandchild, half-aunt/half-uncle of an individual (12.5% of DNA shared on average) |

**Table 2** Degrees of biological relatedness.

### 2.3 Socio-material networks at Çatalhöyük

In previous network studies at Çatalhöyük, the analysis of socio-material networks was used to disentangle patterns of "social" relations using material features as proxies of processes of affiliation, belonging or social co-operation between houses, which represent the smallest unit of analysis (Mazzucato, 2019, 2021; Mazzucato et al., 2022). Following Knappett (2011) these formal network representations are defined as socio-material because they aim to reconstruct social relations using material proxies and they are the result of the interplay of both material and social dynamics (see Mol, 2014). These networks were constructed using individual buildings as nodes and the relationships between nodes were established through the repetition of specific material culture features found within buildings. Buildings have been recognized as suitable units to be used as nodes in the network construction; they are bounded entities, spatially distinguishable from one another and their modular repetition and proximity form the dense agglomeration of the East Mound. Furthermore, houses at Çatalhöyük appear to serve as the primary unit of social organization throughout the entire occupation of the site, despite certain differences and changes through time (Hodder, 2014a, 2022; Hodder and Cessford, 2004; Marciniak, 2019; Marciniak et al., 2015). Buildings at Çatalhöyük, in their completeness, are essential social and ritual units; each of them exhibiting a distinctive and peculiar character (Asouti, 2006; Hodder and Cessford, 2004). In order to have comparable units in the network construction, only buildings that were at least 50% excavated were used (Table S2.2). The majority of buildings utilized in the analyses date to the Early and Middle occupation periods (Table 1). Links between buildings were thus traced through similarities between material culture and through the co-occurrence of specific features. A complex dataset composed of an array of relevant objects and material practices selected from the entire corpus of material culture items excavated by the ÇRP between 1993 and 2017 was used to create bipartite affiliation networks (Table S2.3). This approach, which is based on building connections through material culture co-occurrence, assumes that, at Çatalhöyük where inhabitants adhered to arrays of strong “socio-symbolic codes” through the life of the site, a certain amount of a selected set of material culture homogeneity reflects degrees of social proximity, communal practices or processes of identification or membership of buildings to specific corporate supra-household group (e.g. neighborhood, sodalities). At different scales, material culture has been extensively used to create archaeological similarity networks (e.g. Blake, 2014; Brughmans and Peeples, 2023; Coward, 2010; Donnellan 2020; Jayyab and Gibbon, 2022; Mills et al., 2013; Mol, 2014; Östborn and Gerding, 2014; Pereira at al. 2023).

Socio-material networks were first constructed as bipartite affiliation networks of buildings and material culture and then projected on 1-mode networks of buildings (Mazzucato, 2019, 2021). For the current study, socio-material networks of buildings have been integrated with data related to the genetic relatedness of individuals buried together underneath house floors. In doing so, we aim to gain insight into the relationship between genetic and material culture ties at the site and to identify processes of kinship construction. We utilize a selection of the entire dataset analyzed by the NEOGENE team. The dataset is composed of 81 individuals and of 2,425 pairs of relations between individuals. Genetic kinship estimation was performed using multiple tools as described in (Yüncü et al., 2024), and each pair of genomes was required to share at least 3000 genetic variants (Table S2.1). Within this set of pairs, we selected those with individuals found buried underneath the floor of buildings that were included in the construction of the socio-material network (Table S2.2). Our data allow only inference up to 3rd-degree relatedness, while more distant relatives would be classified as unrelated. We also note that genetic relatedness level estimations are subject to both biological and technical noise. For instance, although 1st-degree kinship estimates are generally reliable at the cutoffs used here, 3rd-degree relatives may still be missed or distant relatives may be inferred as 3rd-degree; this can happen due to the inherent randomness of genetic inheritance as well as the partial nature of ancient genomes (Aktürk et al. 2024).

For this study, the material practices used for generating links between nodes (Table S2.3) are regarded as indicative of forms of relatedness at the site and material similarity links are recognized as concerning both with patterns of belonging or social co-operation between houses and with instances of “material affinities'' between kin components (Goldfarb and Schuster, 2016; Holmes, 2019; Manson, 2008). Previous work highlighted how such material practices define socio-material geographies consisting of different choices at the scale of the individual house and of relations differentially established by social groups to materials, resources and other groups that produced a flexible and resilient social arrangement (Bogaard et al., 2017; Mazzucato, 2019; Mazzucato et al., 2022).

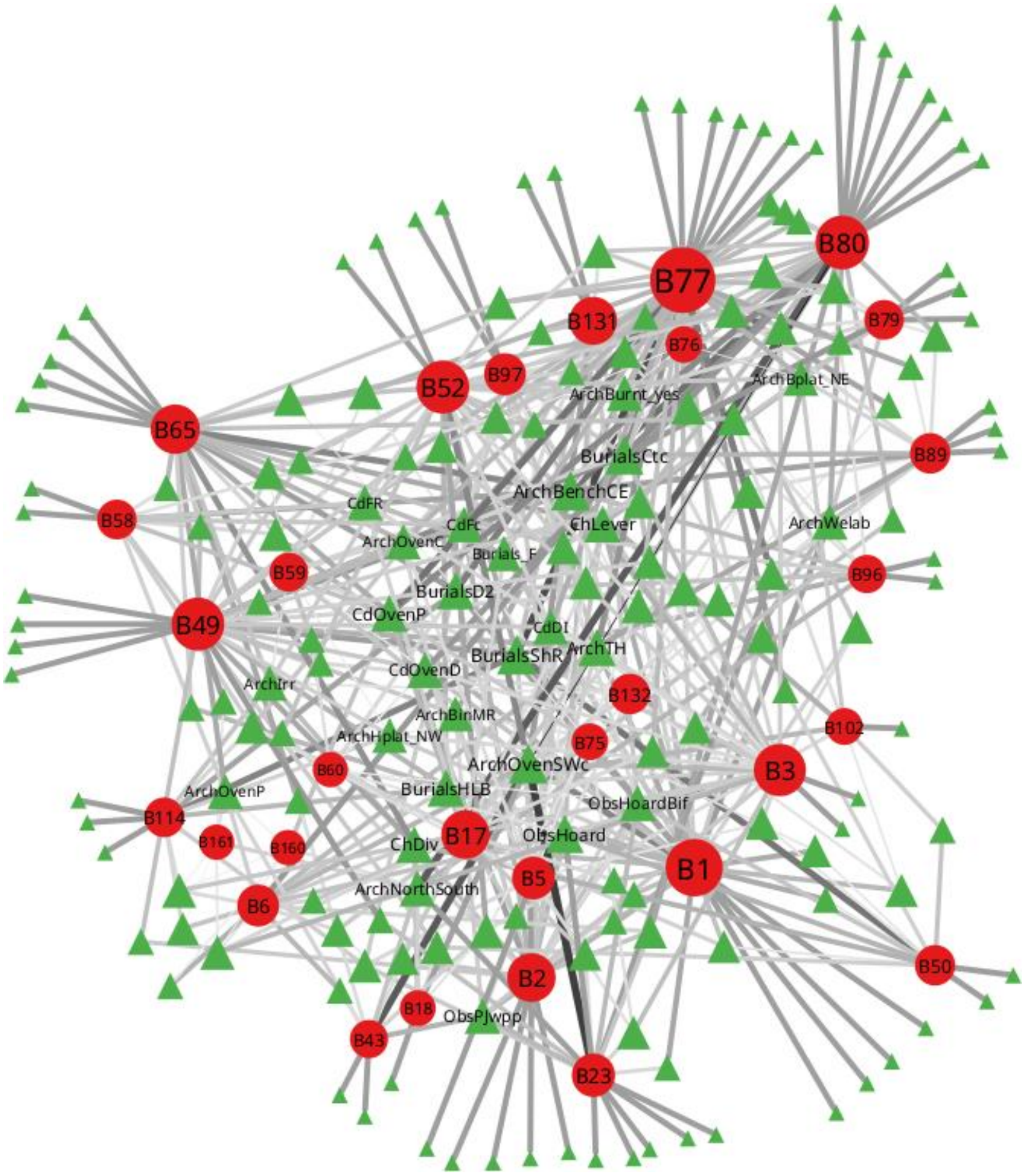

**Figure 2**: Bipartite affiliation network of buildings (red dots) and material culture and practices (green triangles). The color of ties is proportional to the edge betweenness values. High edge betweenness centrality values define edges that connect disconnected parts of the network (see Coscia, 2021 for a definition of edge betweenness).

## 3. Methods and results

### 3.1 Network of material culture and biological relationships

Similarly to the socio-material networks previously constructed (Mazzucato 2019, 2021), the network generated for the current study is created projecting the bipartite affiliation network of buildings and material culture (Figure 2) on a 1-mode network of buildings/nodes and count of common artifacts between pairs of buildings as weighted links between them. For this network, all buildings have been used regardless of their chronological grouping. The chronological "macro-phases" (Early, Middle and Late) used to partition Çatalhöyük's stratigraphy are not useful in the current analysis because it results in the artificial splitting of biological ties that span multiple "macro-phases"; indeed, biological ties may span several generations and crosscut broad chronological phases. Furthermore, relations between buildings and forms of identification and belonging between them endure and are strongly perceived at the site across time. At Çatalhöyük, “temporal depth” is itself a means for creating complex ties between people in relationships of affiliation and dependence (Hodder 2018: 26) and, as Hodder (2018) points out, site stratigraphies should be regarded and investigated as social processes, not only as useful tools “in sorting out chronological sequences” (Hodder 2018: 22).

For this study, statistical significance is used to define the presence/absence of links between buildings. This is done to deal with the densely connected bipartite graph in Figure 2 and to control for the difference in the amount of material culture recovered in buildings; as mentioned earlier, some buildings are much more elaborate and rich in material culture than others (Hodder, 2014b, 2021, 2022; Hodder and Pels, 2010). This difference might result in elaborate buildings sharing a larger number of material items, between themselves or with other less elaborate buildings, solely by chance without the presence of a significant relation. To define statistically significant connections between buildings we use the noise-corrected backboning algorithm which is designed for networks exactly like the one we generated for the current study, where edge weights are determined by discrete counts (Coscia & Neffke, 2017). This approach allows us to correct for the above-mentioned random chance connections due to the difference in elaboration of buildings. This process led to the creation of a weighted network of material culture that has $|V| = 33$ buildings, and $|E| = 81$ significant edges where each building has at least one edge and the network has a single connected component. Edge weight is calculated using the reverse p-value ($1 - p$), since we operate under the assumption that the higher the weight of an edge, the more the two buildings are related. Genetic relatedness is then added to this network as an attribute of each node. All individuals that share a genetic relationship of first, second or third degree, and that are buried underneath the floor of the buildings present in the socio-material network, are included as counts to the properties of each node (Figure 3). Figure 3 shows the socio-material and biological network which is the result of the above-described process. In this network node size corresponds to the number of artifacts recovered per building, the size of edges is defined by the number of shared artifacts between buildings, the edge color is the level of statistical significance and the color of nodes is determined by the number of individuals with genetic relations buried within each building (see table S2.1 for the list of all related individuals). Specifically, in Figure 3 we display the distribution of the 3rd-degree group of genetically related individuals across the socio-material network. From now on, by 2nd-degree or 3rd-degree groups of genetically related individuals, we refer to sets of more than two genetically related individuals that are spread over at least two buildings.

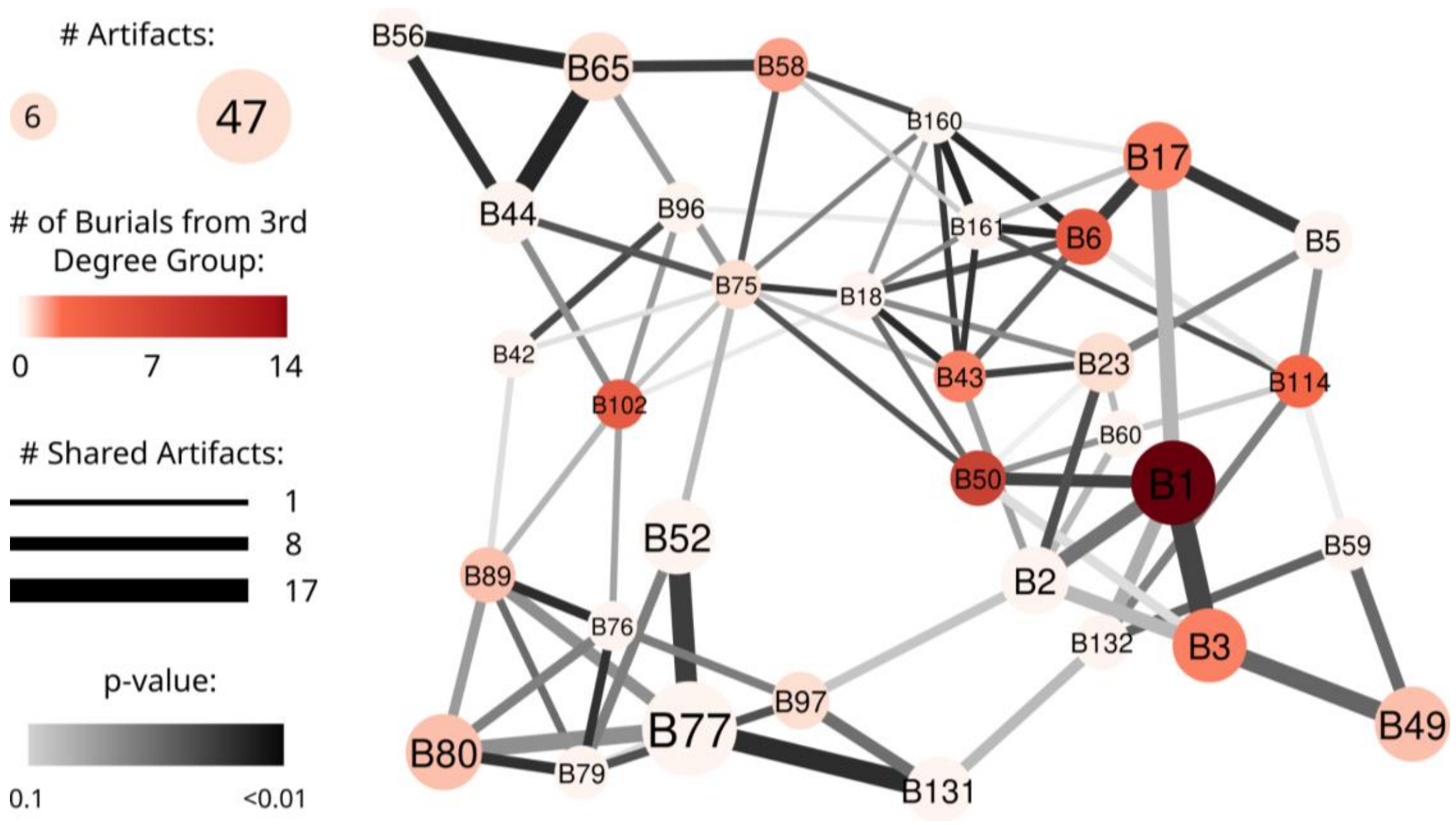

**Figure 3:** Material and biological network. Node size corresponds to the number of artifacts recovered per building, edge size is the number of shared artifacts between buildings, edge color is the level of statistical significance and node color is determined by the number of biologically related individuals within that building (3rd-degree). B.1 is the building with the highest concentration of individuals that share a 3rd-degree relation, 14 individuals, while, for instance, B.49 has only two 3$^{rd}$-degree related individuals and B.160 none.

Intuitively, this seems to concentrate in the network – 26 out of its 64 members are buried in the B.1, B.3, and B.50 buildings, which are all connected to each other. However, to determine whether this observation is statistically significant we need to quantify it.

### 3.2 Network dispersion of groups of genetically related individuals

In statistics, variance is a measure of dispersion: it measures how much a numerical vector distributes away from its average value. Recently, researchers extended the concept of variance to be applicable to the case in which the numerical vector is distributed over a network (Devriendt et al., 2022; Coscia, 2020); for a review of methods see (Coscia et al., 2020). A vector has low variance if it concentrates in the network while it has a high variance if it is dispersed across nodes that are far away in the structure. We have the minimum possible network variance value when a single node has a non-zero value and all other nodes have zero value, i.e. the vector is concentrated in a single node.

To calculate the variance of a group of related individuals we need to represent it as a vector. The vector has one numerical entry per building, equal to the count of group members buried in that building – e.g.

for the group in Figure 3, B.1 is equal to 14, B.3 is equal to 4, and so on. Buildings with no individuals from the 2nd or 3rd-degree groups are assigned a value of 0.

It should be stressed that, Yüncü and colleagues (2024) couldn't identify any 1st-degree relatives between burials in different buildings; therefore 1st-degree relations cannot be included in the analysis of variance and we are limited to 2nd- and 3rd-degree relations.

Applying the network variance calculation to the vector and network constructed, results in the network variance values we report in Table 4. Taken in isolation, these numbers are hard to interpret. To contextualize them, we must compare them with a null model.

| Groups | Degree | Size (# of individuals) | Variance |
|---|---|---|---|
| 1 | 2nd | 8 | 0.031 |
| 2 | 3rd | 64 | 0.155 |

**Table 4:** The size (number of individuals with genetic kin) and network variance in the material culture network of all non-trivial groups of genetically related individuals (more than two members spanning at least two buildings). Note that group #2 contains group #1.

### 3.3 Null model

To understand whether 2nd- or 3rd-degree groups are more or less concentrated than expected in the network, we can compare them with a null model. In the null hypothesis, the location of group members is unrelated to the building relationships as inferred from the material culture records. This means that we can create a "null group" by simply extracting random buildings from the building list. For instance, since Group #1 is composed of eight members, we extract eight random buildings. To be consistent with our representation of a group as a numerical vector, we will perform an extraction with replacement: several null group members can be buried in the same building, just like members of Group #1 could. We also weigh the extraction probability with the number of burials that were effectively found in the buildings – so that a building with more burials is more likely to be extracted in the null model.

To obtain a statistically robust result, we generate 100,000 null groups and we record the network variance value for each of them. For 3rd-degree groups, we average their network variance values and compare them with the average of their null versions taken together.

Figure 4 compares the observed value from the real groups of genetically related individuals to the distributions of their null versions. For both the 2nd-degree group (Figure 4(a)) and the 3rd-degree one (Figure 4(b)) we see that fewer than 0.1% of null models score a network variance value lower than or equal to the one we observe. This corresponds to a pseudo $p < 0.001$, suggesting a highly statistically significant result.

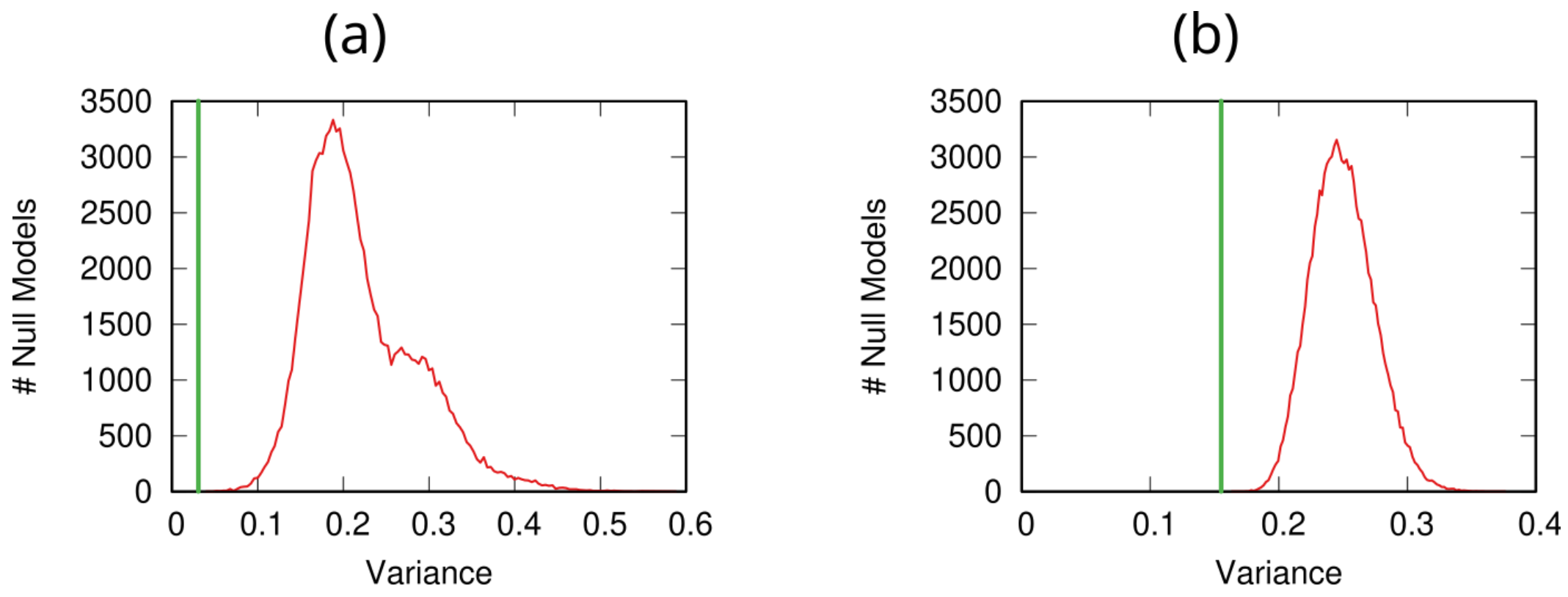

**Figure 4**: Comparing the observed network variance with the expectation from a null model in the material culture network. The red line shows the number of null models (y axis) with a given value of network variance (x axis). The vertical green line shows the observed value. (a) 2nd-degree version. (b) 3rd-degree version.

From this analysis we can conclude that **material culture similarity between buildings indeed gives valuable information about potential biological relationships between individuals**. The observed groups of biologically related individuals concentrate more in the network than we would expect from random chance. Therefore, being close within the material culture network increases the likelihood of two buildings to include biologically related individuals buried underneath their floors.

### 3.4 Material culture, spatial and temporal distances

In this section we evaluate the role of spatial proximity and temporal variation in defining the relationship between genetic links and material culture. We do so first by constructing a new network where links between pairs of buildings are determined exclusively by Euclidean and temporal distance (geotemporal network) and then, by performing regression using different models. Furthermore, in this section, we want to disentangle space and time effects from the relationship between material culture similarities and genetic relatedness. To do so we calculate the regression residual network and we measure the variance of biological relatedness between individuals on the network.

We know from previous network studies at Çatalhöyük, that geographical proximity is an important social organizing principle at the site. Quadratic assignment procedure (QAP) correlation and regression (MR-QAP) are used as tools for testing the hypothesis of the role of Euclidean distance in forming ties between pairs of nodes (Mazzucato, 2021). These methods demonstrate that, mainly in the Early and Middle periods, the location of buildings on the East Mound define patterns of socio-material affiliation; buildings that were close to each other were more likely to be similar in terms of their material culture assemblages (Mazzucato, 2021). This observation appears to substantiate the hypothesis that social associations larger than the individual building materialized in clusters of adjacent buildings that were sharing roofs, similar to what has been observed at other Anatolian sites (e.g. Canhasan III and Aşıklı Höyük) (Düring and Marciniak, 2006; Düring, 2007). The same observation is true for sequences of overlapping buildings that appear to be deeply entangled in material relations; buildings constructed in the same location are more likely to be similar in terms of material practices (Mazzucato, 2021). However, previous socio-material network analyses revealed that geographical proximity accounts for only a part of the behavior of the dataset, and some of the ties between buildings crosscut their geographical location (Mazzucato, 2019, 2021, 2022). Indeed, the entire community of Çatalhöyük maintained a distinctive identity throughout its life by means of what appear to be strong socio-religious rules (Hodder 2021). The long-distance ties were probably of major importance in maintaining the coherent and distinctive identity of the Çatalhöyük community, which, as Baird (2019) suggests, was very likely essential for the long-term existence of these communities.

### 3.4.1 Geotemporal Network

The geotemporal network is constructed using only space and time to define the links between buildings. For each building pair, we calculate the Euclidean distance between them using the *X*, *Y*, and *Z* coordinates. We can obtain the temporal dimension (*Z*) from the level in which a building was excavated. We need to convert the chronological levels to a numerical value and we decide to make *Z* span the same range size as *X* and *Y*, not to make any dimension dominate the result in the distance calculation. We distribute layers uniformly in this range, in chronological order.

Once we estimate the distance between each pair of buildings, we select the closest pairs as the edges of the network so that the following conditions are satisfied. First, each building needs to have at least one connection to another building. Second, there cannot be nodes without paths to the rest of the network – i.e. part of disconnected components. These are necessary conditions to calculate network variance. Finally, the network should have the same number of edges as the material culture network. To satisfy the first two conditions, we need to replace the longest 4 out of the 81 shortest edges with longer edges that can keep the network as a single connected component. Figure 5 shows the resulting network. We used the inverse of the geotemporal distance as edge weights, since we want them to describe the strength of a relationship between two buildings – and thus the weight needs to represent proximity, not distance. We then calculated the network variance for all the groups of 2nd and 3rd-degree relatives; Table 5 contains the network variance values.

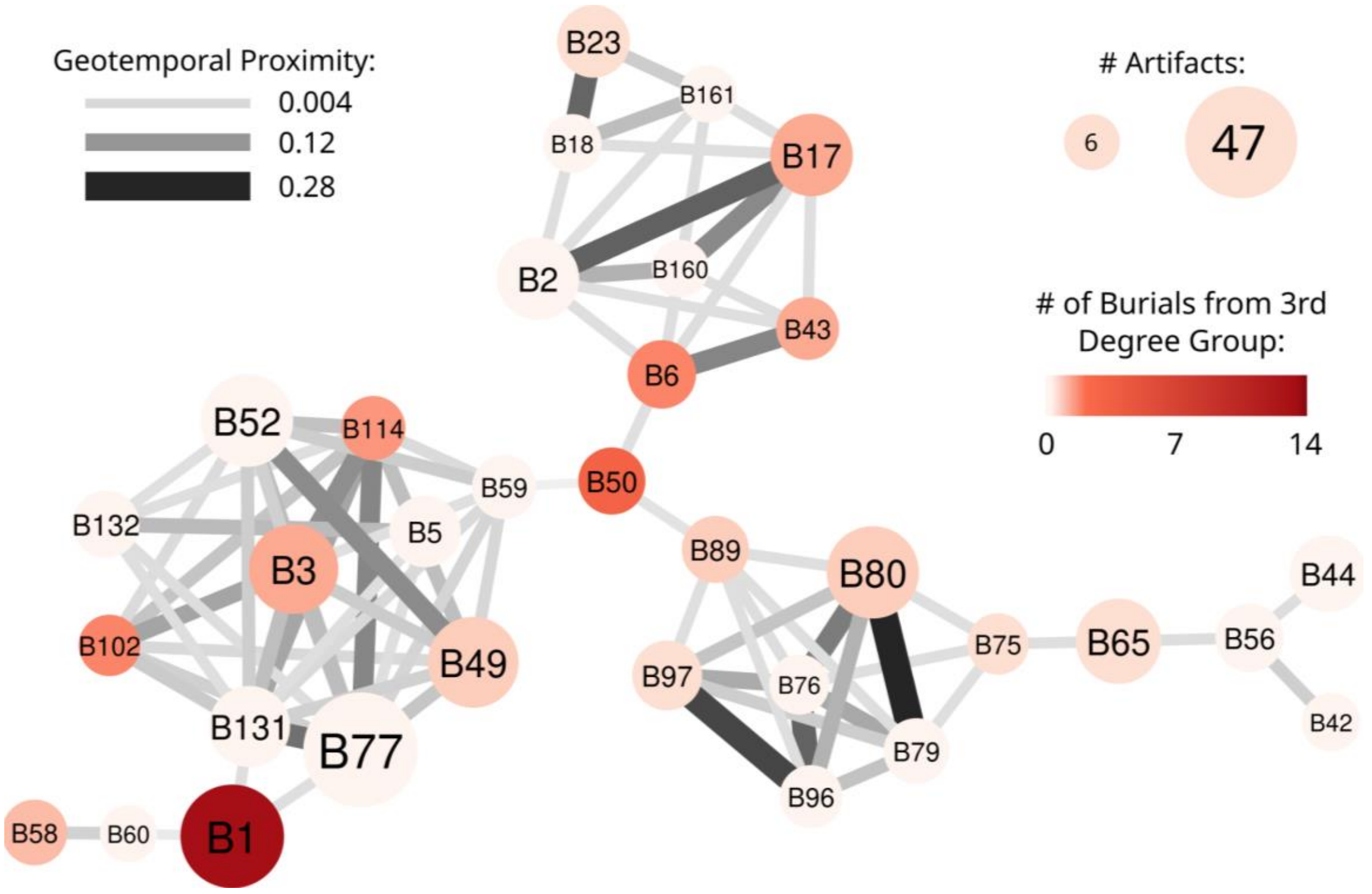

**Figure 5:** The geotemporal building network. Node size and color consistent with Figure 3. Edge size and color is proportional to geotemporal proximity.

| Group | Degree | Size | Variance |
|---|---|---|---|
| 1 | 2nd | 8 | 0.004 |
| 2 | 3rd | 64 | 1.968 |

**Table 5**: The size (in number of members) and network variance in the geotemporal network of all non-trivial groups of genetically related individuals (more than two members spanning at least two buildings). Note that group #2 contains group #1.

We completed the analysis by implementing the null model that we had applied to the dataset in the previous section; this permitted us to observe that for the group of 2nd-degree related individuals, the concentration is higher than expected ($p < 0.01$). Regarding the dispersion of the group of 3rd-degree relatives, we obtained a result consistent with the null hypothesis ($p > 0.1$). Which means that individuals that share a 3rd-degree genetic relations don't concentrate spatially on the mound. Figures 6(a) and 6(b) compare the null model value distributions with the observed value in the groups of 2nd- and 3rd-degree genetically related individual, respectively. The multimodal distribution of values from the null model is a result of the spatio-temporal organization of the settlement, with clear divisions between north and south parts.

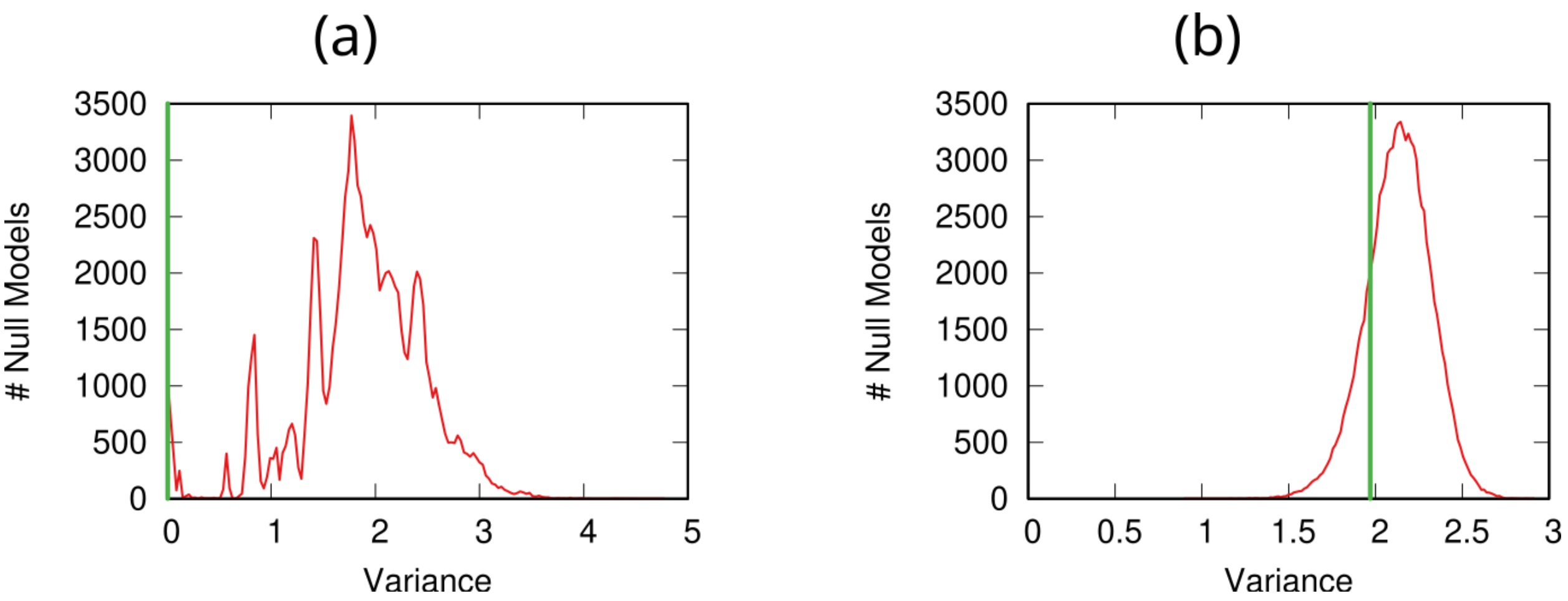

**Figure 6:** Comparing the observed network variance with the expectation from a null model in the geotemporal network. Same legend as Figure 2. (a) 2nd-degree version. (b) 3rd-degree version.

In summary, the geotemporal factor is related to the dispersion of the groups across buildings: groups of genetically related individuals concentrate more than we would expect from random chance, but only for strong genetic ties (2nd-degree). This implies that close biological relatedness concentrates spatially (and temporally), while weaker ties do not. While the chronological concentration is to be expected, the spatial concentration appears to be more interesting. With the intention of further investigating the relations between space, time and similarities in material culture and to further determine how such variables interact with biological relatedness on site, we construct the material culture residual network. This gives us the opportunity to observe how biological relatedness behaves on the network of material culture similarities after removing the space and time effect.

#### 3.4.2 Material culture residual network and network dispersion of groups of genetically related individuals

In this section we want to disentangle the effect of space and time variables on material culture with the aim of establishing if similarities in material culture provide information regarding the distribution of biological relatedness beyond geotemporal proximity. To this end we performed an Ordinary Least Squares (OLS) regression using models 1, 2, and 3 (S1.3), where space (model 1: Geo.Dist.) and time (model 2: Temp.Dist.) are considered individually and then together but as separate entities (model 3). If we run such a model, we obtain the expected negative and significant coefficients for both distance measures: the further two buildings are in space (or in time) the less material culture they share, keeping the other distance constant. Table 6 shows the results of the regression (see S1.1 for the regression formulas).

| | **Dependent variable:** | | |
|---|---|---|---|
| | **p-value (Material Culture)** | | |
| | -1 | -2 | -3 |
| Geo. Dist. | -0.050$^{***}$ | | -0.041$^{***}$ |
| | (0.015) | | (0.015) |
| Temp. Dist. | | -0.068$^{***}$ | -0.063$^{***}$ |
| | | (0.015) | (0.015) |
| Constant | 2.044$^{***}$ | 2.107$^{***}$ | 2.250$^{***}$ |
| | (0.067) | (0.063) | (0.083) |
| | | | |
| Observations | 528 | 528 | 528 |
| $R^2$ | 0.019 | 0.037 | 0.050 |
| Adjusted $R^2$ | 0.018 | 0.035 | 0.046 |
| Residual Std. Error | 0.551 (df = 526) | 0.546 (df = 526) | 0.543 (df = 525) |
| F Statistic | 10.446$^{***}$ (df = 1; 526) | 20.209$^{***}$ (df = 1; 526) | 13.809$^{***}$ (df = 2; 525) |
| *Note:* | $^{*}$p<0.1; $^{**}$p<0.05; $^{***}$p<0.01 | | |

**Table 6:** The results of the OLS between material culture and geographical (1. Geo.Dist.) temporal distance (2. Temp.Dist) and both variables (3.). Top section: coefficient on top and its standard error in parenthesis below

(Geo/Temp.Dist. row); coefficient and its standard error (Constant row). Bottom section: summary statistics of the regression.

After performing the regression, we now want to investigate how biological relatedness relates to material culture only, without the geospatial effect. To do this we use the residuals of the regression model (model 3) as the edge weights of a new network, the material culture residual network (S1.4). Figure 7 depicts the residual network. It is harder to get an intuitive reading of the picture than it was for the previous two illustrations. The figure shows those material culture relationships between buildings that exceed our expectation given the buildings' geotemporal location.

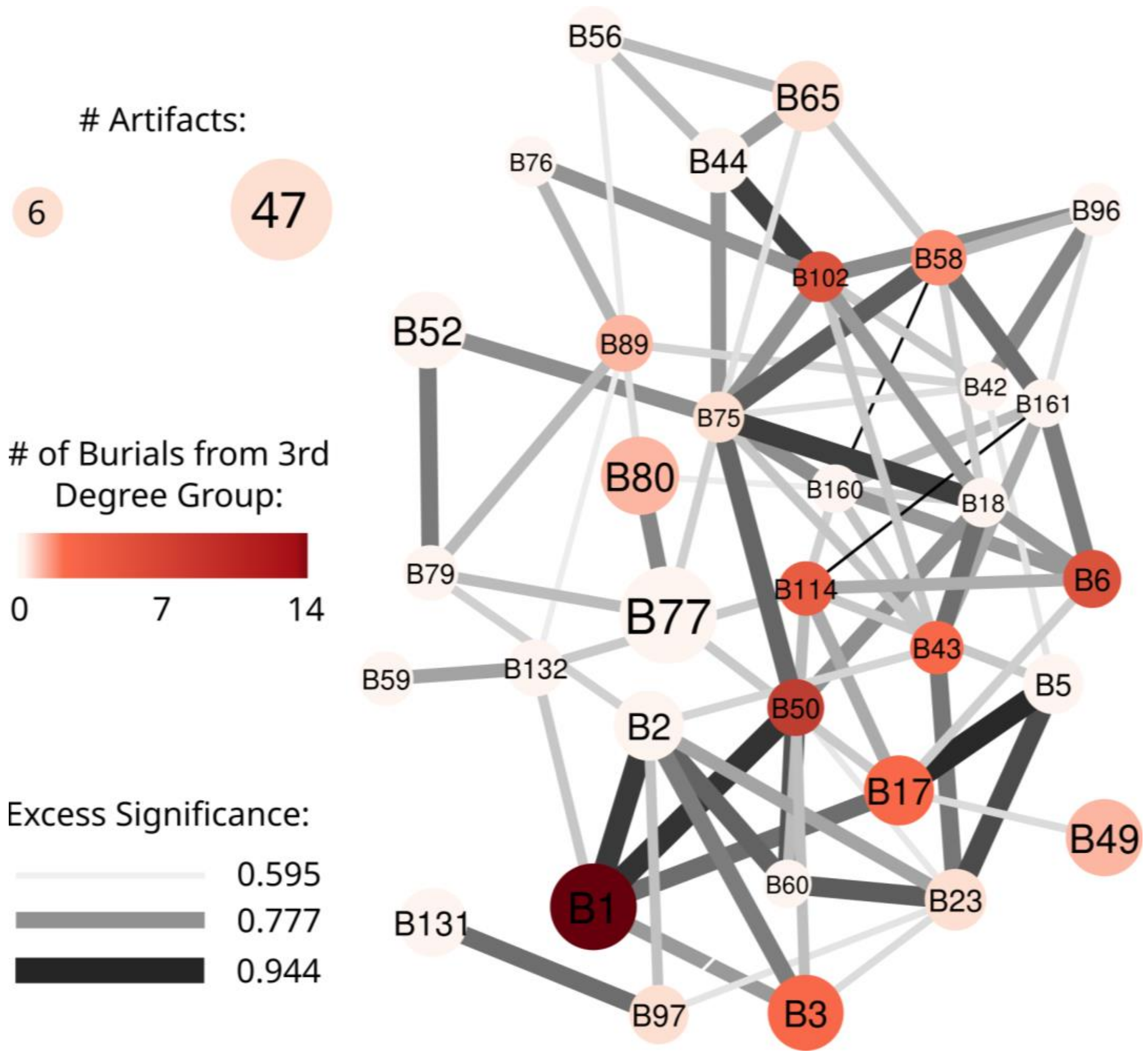

**Figure 7:** The residual network. Node size and color consistent with Figure 1. Edge size and color is proportional to the excess material culture significance.

For instance, consider buildings B.79 and B.80. In Figure 1 we see that they share a significant amount of artifacts – they are connected with a dark link. However, from Figure 3 we see that they are also extremely close to each other. This results in them not being connected in Figure 5 because, while their material culture relationship is highly significant, it does not exceed our expectation given their proximity by a large enough margin to be included in this network.

We repeat the network variance experiment, this time on the extracted residual networks. Similarly to the previous tests, we use the weighted network, both for the groups of 2nd and 3rd degree relatives. In the residual network, each edge's weight is the residual inverted p-value we obtain from the regression. To interpret the variance values, we need to compare again against the distribution of variances we get from randomized versions of the groups of genetically related individuals.

Figure 8 shows the resulting distributions, confirming visually that the expected variance from the observed group (in green) is lower than the expected variance from the null model. We confirm that for both the 2nd-degree group (Figure 9(a)) and the 3rd-degree group (Figure 9(b)), we observe a higher concentration than expected ($p < 0.01$).

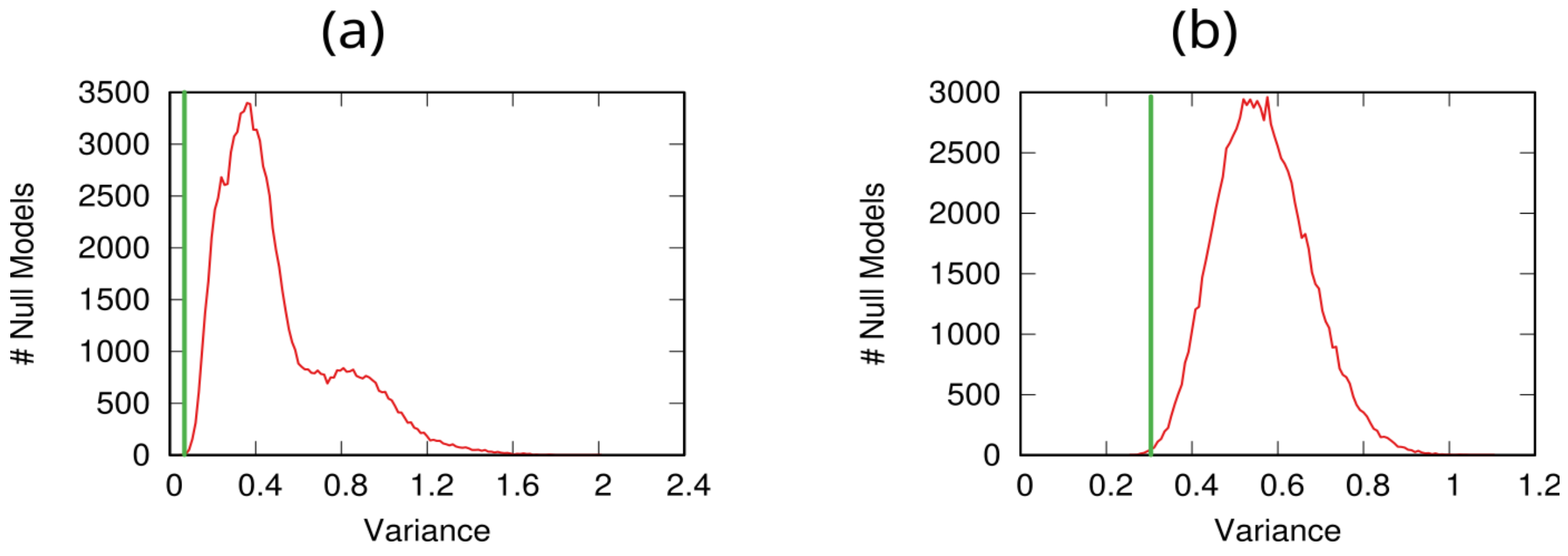

**Figure 8:** Comparing the observed network variance (green) with the expectation from a null model (red) in the residual network., (a) 2nd degree version. (b) 3rd degree version.

Furthermore, we now investigate what would happen if we were to ignore the temporal dimension and we would concentrate only on the spatial dimension. Thus, instead of using the residuals of model 3 (space and time) we use the residuals from model 1 and, therefore, we keep the spatial effect and remove the temporal one. Figure 9 shows the results, which are again in perfect agreement with all tests run on any version of the residual network, showing high concentration with $p < 0.01$.

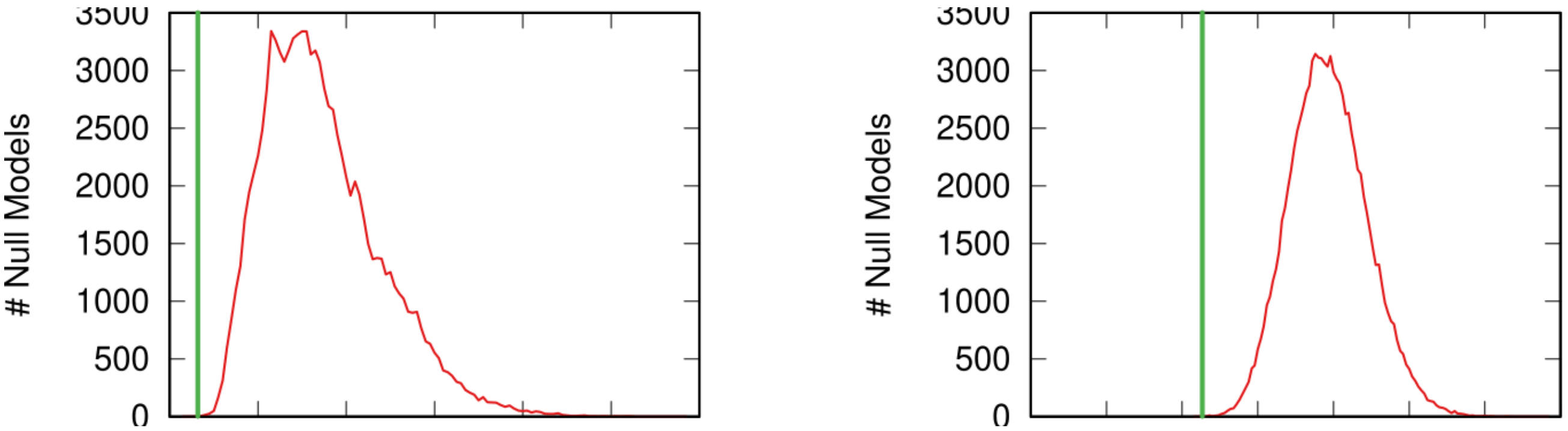

Figure 9: Comparing the observed network variance (green) with the expectation from a null model (red) in the material culture residual network of the spatial-only network. Same legend as Figure 2. (a) 2nd degree version. (b) 3rd degree version.

In conclusion, this section shows that the groups of 2nd- and 3rd-degree genetic relatives concentrate more than expected in the residual network. **This means that material culture provides valuable information about biological relationships even after we take out from it the effect of geotemporal proximity**. Furthermore, **we observe that when we use the material culture residual network of the spatial-only network both 2nd and 3rd degree relations concentrate,** and this furthermore highlights the importance of spatial proximity on site.

## 4. Discussion and conclusions

Network methods allow us to explore the relationship between individuals' biological relatedness and material culture similarities at Çatalhöyük. Additionally, network analyses were used to investigate how biological relations are distributed spatially between buildings. By calculating the network variance of the groups of 2nd and 3rd-degree genetic relatives, we see that biological relatedness correlates with material culture similarities and concentrates in space.

What do these findings tell us about the modes of relatedness of this large Neolithic site? Can this observation provide insights into the way people at Çatalhöyük constructed their concepts of relatedness? The analyses suggested a spatial (geographical) concentration of genetic relations on the network of buildings; while we could not use the 1st-degree relations between individuals in our analyses, because they were found only located within buildings, both 2nd-degree and 3rd-degree biological relations appear to be more geographically concentrated on the East Mound than we would expect by chance (Figure 6 and 9). It should be stressed that 2nd-degree relations are found in our dataset restricted to a few buildings in a small part of the South Area, however, 3rd-degree connections, which are geographically distributed in both the South and North Areas, appear to be equally more spatially concentrated than expected. Yüncü and colleagues (2024) noticed a similar pattern for all biological connections which tend to be concentrated either inside the same building, or between buildings that overlapped each other, or between buildings in proximity. Furthermore, it interesting to notice that the group of 3rd-degree genetic relatives does not concentrate in the geotemporal network when we take into account both time and space together (Figure 6). On the contrary, it concentrates in space when we remove time from the model (Figure 9).

As previously highlighted, a previous genetic study based on small samples sizes had suggested that in the 7$^{th}$ millennium BCE at sites like Çatalhöyük and Barcın genetic related individuals buried in the same building are rare and that biological relatedness appears to play a marginal role in the selection of individuals to be buried underneath the floor of the same house (Yaka et al., 2021). Recent archaeogenomic research shows that, while this observation holds true for the Late Period at Çatalhöyük, the Early Period shows a different pattern, with genetically related individuals buried more often within the same building, even if genetic kin is still not the sole criteria used to determine the selection of individuals for co-burial (Yüncü et al., 2024). Despite this change through time, we found that genetic relatedness between individuals across buildings persists at specific localities throughout the entire sequence. The closer in space two buildings are, the more likely it is that they contain genetically related individuals. This finding mirrors an array of observations made across a variety of material categories, suggesting a crucial role of spatial proximity as a principle of social organization at the site (e.g. Bogaard et al., 2017; Hodder, 2013, 2014a, 2022; Tung, 2013; Yalman et al., 2013,). The socio-material network analyses previously performed on material similarities alone revealed that similarities in material culture across buildings tend to cluster spatially suggesting possible sets of affiliations between spatially proximate buildings that formed social groups larger than the house, at least for the Early and Middle levels (Table 1) (Mazzucato 2019, 2021; Mazzucato et al., 2022). As previously mentioned, many scholars acknowledge the importance of the affiliations between groups of adjacent buildings at the site. The observed clustering of genetic relations between individuals across spatially adjacent buildings further corroborates the importance of location for understanding how relatedness at different scales unfolded at the site. The role played by building location at Çatalhöyük has also been observed recently using isotopic analyses of human diet (Pearson et al., 2021, 2023). Pearson and colleagues discovered that, when the diet signature of individuals buried under the floor of adjacent houses is examined, a statistically significant difference is noticed between clusters of spatially proximate buildings regardless of their chronological relationship. When the same statistical analyses are applied to nearby buildings arranged chronologically by levels the results are not statistically significant. We noticed a similar pattern when we analyzed the 3rd-degree biological connections across buildings which cluster in space only when time is removed from the model (Figure6 and 9). In a similar vein, recent archaeobotanical research has further highlighted similarities in land use by neighboring buildings or by buildings in the same sector of the East Mound that are observed across time periods (Bogaard et al., 2017, 2021). Considering the recent archaeobotanical evidence, a model of land management arranged in radial slices of arable land around the site has been proposed. According to this model, spatially proximate buildings on site were exploiting the same "wedge" of fields radiating from the mound into the surrounding landscape. Such strips of land were, therefore, made of fields that were close to the settlement and that could be intensively cultivated, and fields distant from the settlement where cultivation intensity decreased (Bogaard et al. 2021). This model of land management acknowledges the role of the location of buildings on the mound and the relationship with a specific part of the landscape that such location implies. The location of buildings in a specific sector of the East Mound may have entailed a complex set of relationships with the landscape and with people, things and localities within and beyond the site (at different scales). The way ideas of relatedness were constructed and developed through time may also have included this complex relationship with the landscape. Ethnographers working in Melanesia observe that "kinship is landscape" (Stasch, 2009; Telban, 2019:488). James Leach, who studies the "modes of relatedness" among the people of the Rai Coast of Papua New Guinea, stresses how land is an active force in social processes for the people living in the area. In the Rai Coast, land produces the very substance that humans share by means of the consumption of food from the same place (Leach, 2003, 2019). The relationship between people and the landscape should be understood as a constitutive one, as a relation that brings people into life and defines kinship bonds between people (Leach, 2003, 2009). In a similar vein, Sandra Bramford describes the role that land has in producing ties between generations within the Kamea people of Papua New Guinea. For the Kamea people, land is regarded as the way of constructing social relationships through time and as the determinant of intergenerational ties (Bamford, 2007, 2009). It is the relationship with non-human resources and the environment that grants young people their place in the world and in the network of "remembered social relationships'' and kinship links (Bamford, 2007:56). At Çatalhöyük, complex networks of ''remembered social relationships'' define the fabric of the site and its social geography at different scales by means of a strong commitment to specific localities and through a process of cyclical returning to "persistent places" on the Mound and, perhaps, in the broader landscape. This commitment to localities in the landscape, to their resources and people must have been deeply entangled with the process of constructing concepts of relatedness and kinship bonds at Çatalhöyük. These practices weave together biogenetic proximity, sociocultural processes and the relationship with non-human entities through a process that does not distinguish between biology and social dynamics, nor between human and non-human entities. Based on recent genetic results, maternal lineages appear to have played a crucial role in situating these kin groups within specific localities (Yüncü et al., 2024).

In a recent article, Hadad (2024) stresses the importance of the “rhythmical connection to a place” (Hadad, 2024:13) as the defining quality of the development of Neolithic sedentism, which was deeply entangled with the very nature of the building material at Çatalhöyük and the need for repetitive reconstruction, replastering and repairing of mudbrick houses. Morton (2007) notes how in Tswana (northern Botswana) the “building over time” of houses and the practices of repair “prompt genealogical remembering” and that this is an essential element in “evoke and sustain kin relations” (Morton, 2007:159). The practice of placing human and animal remains within the fabric of the house was likely part of the “kinwork” of forging, sustaining and recognizing complex kin relations which included non-human entities. Aurochsen remains were frequently incorporated in the house and they were likely recognized as ancestors and important kin members (Haddow at al., 2016; Russell, 2022; Twiss and Russell 2009).

The flexible and situated nature of kinship relations is evident in the changes through time observed at Çatalhöyük by Yüncü and colleagues (2024). In this regard, it is interesting that changes in the pattern of co-burial relatedness from the Early period to the Late period developed alongside crucial shifts in the way the inhabitants of the site used the landscape and its resources (e.g. Hodder, 2014a, 2021; Marciniak et al. 2015a). It is likely that these changes entailed modifications in patterns of relationships/alliances with other groups in the Konya Plain and beyond; Russell (2022) suggests that the late appearance of domestic cattle in the Late levels at Çatalhöyük may have been correlated with changes in kinship patterns. The social relations afforded by wild cattle and the practice of hunting them in the landscape were very likely completely different from the one afforded by domestic cattle and the different relations with landscape resources (Russell, 2022).

But this is not the only aspect highlighted by the network analyses and spatial proximity is not the only factor driving the social organization of the settlement (Hodder 2022; Mazzucato, 2019, 2021). Indeed, network variance suggests that similarities of material choices between buildings represent a predictor of the likelihood of finding biologically related individuals buried in those buildings. When the effect of space and time is removed from the network of material culture, the measure of dispersion of the groups of genetically related individuals persists as lower than expected by chance (Figure 8). Indeed, material culture similarities appear to be the best predictor of the presence of biologically related individuals buried in buildings beyond their spatial proximity. This appears to mirror the results of previous socio-material network analyses which, on the basis of material relations alone, identified ties that crosscut the spatial location of buildings (Mazzucato, 2019, 2021; Mazzucato et al., 2022). As Hodder observes, “neighborhoods were probably cross-cut by co-burying and co-eating groups” (Hodder 2022:13).

How can the correlation between material cultural choices and biological relatedness be interpreted and what does it say of the way concepts of relatedness were constructed at Çatalhöyük?

On a different scale, the ways in which material culture similarities and aDNA data have been integrated within archaeological discourse has been largely problematized. Scholars point to the widespread oversimplistic interpretation of the complexity of archaeogenomic datasets, which have often been seen through the lens of an outdated culture-historical paradigm (Furholt, 2018, 2019a, 2019b, 2021). This antiquated perspective sees prehistoric societal dynamics as a process of interaction between fundamentally bounded archaeological cultures which are recognized on the basis of shared material culture traits and are regarded as homogeneous social groups and genetic populations (see Wunderlich et al., 2023 for a recent review).

At an intra-site scale, the similarities in material culture and the assemblages of co-occurring practices are instead seen to represent different choices at the scale of the individual house; they are probably the result of relations differentially established by social groups to materials, resources and other groups (e.g. Bogaard et al. 2017; Hodder and Tsoraki 2023). Additionally, the concept of “material affinities” offers a useful framework to think about these similarities and their correlation with biological relatedness at a micro-scale (Holmes 2019:174). This concept is used to explore the many ways objects and practices are *passed on* along networks of relatedness in time and space. Within this framework, materiality is recognized as a fundamental aspect of doing “kinwork” (i.e. the practice of creating and sustaining kinship bonds) (Holmes, 2019; Johnston, 2021; Manson, 2008). Holmes (2019) describes the idea of the *passing on* of objects and practices as an umbrella term that comprises a variety of processes which entails, for example, the sharing of similar practices because they “hold some sentimental significance” or the circulation of heirlooms and “family reminders” (Holmes, 2019:175, 185). She also stresses that *passing on* is not only symbolic or sentimental but it is related to the “materiality of objects” and their affordances. The “material affinities” between buildings at Çatalhöyük may mirror these processes of “doing relatedness” and “kinwork” within an intimate scale. Indeed, kinship bonds are forged through material connections and objects constitute the material media through which kinship relations are experienced and performed across generations through their material biographies (Abram and Lien, 2023; Goldfarb and Shuster, 2016; Johnston, 2021; Lien and Abram, 2023; Weiner, 1992). At Çatalhöyük, biological relatedness is a part of kin-making practices which are never fixed, but always creative and negotiated.